\documentclass[aps,pra,twocolumn]{revtex4-1}
\usepackage{color,amsthm,amsmath,amsfonts,graphicx,bm,epsfig,float,siunitx,multirow,xr,enumerate,epstopdf}
\usepackage[breaklinks=true,colorlinks=true,linkcolor=blue,urlcolor=blue,citecolor=blue]{hyperref}
\usepackage[T1]{fontenc}

\begin{document}

\title{Nuclear Spin Squeezing Based on Spin-Exchange Collisions}
\author{He-bin Zhang$^{1, 2}$}
\author{Yuanjiang Tang$^{2}$}
\author{Yong-Chun Liu$^{2, 3}$}
\email{ycliu@tsinghua.edu.cn}
\affiliation{$^{1}$School of Artificial Intelligence, Hubei Polytechnic University, Huangshi 435003, P. R. China}
\affiliation{$^{2}$State Key Laboratory of Low-Dimensional Quantum Physics, Department of Physics, Tsinghua University, Beijing 100084, P. R. China}
\affiliation{$^{3}$Frontier Science Center for Quantum Information, Beijing 100084, P. R. China}
\date{\today }

\begin{abstract}
Isolation from environment leads to the days-long lifetime of noble-gas nuclear spins, but also brings great challenges to the preparation, manipulation, and measurement of nuclear-spin quantum states. 
Here we find that nuclear spin squeezing, with ultra-long lifetime and huge atomic number, can be efficiently obtained and manipulated based on the coherent spin-exchange interaction between alkali-metal and noble-gas ensembles. 
Thanks to the considerable advantage of our proposal in preparing the spin squeezing of the macroscopic atomic ensemble with a huge atomic number, even the nuclear spin-squeezed state containing $10^{20}$ atoms or more is obtainable with preexisting techniques.
Further, the days-long storage and measurement of nuclear spin squeezing can be performed by the coherent manipulation of a magnetic field.  
This proposal can be implemented in hot atomic ensembles, whose ease of access and high adaptability to various environments will significantly facilitate the research and application of nuclear-spin nonclassical states in precision measurement, quantum information, and fundamental physics. 

\end{abstract}

%\keywords{}

\maketitle


\section{Introduction}

The spin-squeezed state (SSS), similar to the squeezed state of light, is a quantum entangled state of collective spins, and can reduce quantum noise below the standard quantum limit. Since being proposed~\cite{Kitagawa1993_Squeezed, Wineland1992_Spin, Wineland1994_Squeezed}, SSS has attracted considerable attention theoretically and experimentally, due to the important applications in various fields, such as high-precision measurement~\cite{Wineland1992_Spin, Polzik2008_The, Gross2010_Nonlinear, Riedel2010_Atom-chip-based, Bao2020_Spin}, entanglement detection~\cite{Guhne2009_Entanglement}, quantum information~\cite{Sorensen2001_Many-particle, Korbicz2005_Spin}, and foundations of quantum physics~\cite{Ma2011_Quantum}. 
Several different methods for the generation of this quantum state have been developed, including squeezing transfer from light to atoms~\cite{Kuzmich1997_Spin, Hald1999_Spin, Schori2002_Recording}, QND method~\cite{Kuzmich2000_Generation, Bao2020_Spin, Kong2020_Measurement-induced, Kuzmich1999_Quantum, Takano2009_Spin, Auzinsh2004_Can, Kuzmich2004_Nonsymmetric, Hosten2016_Measurement}, squeezing interaction with nonlinear twisting Hamiltonians~\cite{Kitagawa1993_Squeezed, Chaudhury2007_Quantum, Gross2010_Nonlinear, Riedel2010_Atom-chip-based, Liu2011_Spin, Fernholz2008_Spin, Chen2019_Extreme}. 

Meanwhile, the holding time of quantum states, determined by the lifetime of quantum objects, is of crucial importance in quantum science and technology. Due to the complete electronic shell, the non-zero nuclear spins in rare isotopes of noble gases, e.g., ${}^3$He and ${}^{129}$Xe, are well isolated from environment. Therefore, the nuclear-spin ensemble of these noble gases features extraordinary long lifetime of several days at room temperature and is the macroscopic quantum system with the longest lifetime currently known~\cite{Gentile2017_Optically}. 
Currently, noble-gas nuclear spins have been widely employed in precision measurement~\cite{Kornack2005_Nuclear, Walker2016_Spin-Exchange-Pumped, Heil2017_Helium, Jiang2022_Floquet}, medical imaging ~\cite{Albert1994_Biological, Middleton1995_MR, Couch2015_Hyperpolarized}, and searches for new physics~\cite{Lee2018_Improved, Safronova2018_Search, Chupp2019_Electric, Jiang2021_Search}. 
However, these studies are only based on the polarization of nuclear spins without access to nonclassical properties, due to the fact that these nuclear spins are optically inaccessible and can only be polarized by indirect optical pumping methods, including metastability-exchange collisions~\cite{Gentile2017_Optically, Batz2011_Fundamentals}, and spin-exchange collisions between alkali-metal and noble-gas spins~\cite{Gentile2017_Optically, Walker1997_Spin-exchange, Bhaskar1982_Efficiency, Bouchiat1960_Nuclear}. Although some pioneering schemes on noble-gas nuclear spin squeezing have been proposed based on metastability-exchange collisions in ${}^3$He ensemble coupled with optical cavity~\cite{Dantan2005_Long-Lived, Serafin2021_Nuclear}, the incoherent nature of metastability-exchange collisions and the corresponding manipulation inevitably imposes limitations on the study and application of nuclear-spin quantum states. 
Remarkably, it has recently been demonstrated in atomic vapors that a strong coherent coupling between collective alkali-metal spins and noble-gas nuclear spins can be established based on spin-exchange collisions~\cite{Katz2020_Long-Lived, Shaham2022_Strong, Katz2022_Quantum, Katz2022_Optical}. This interaction opens the door to coherent manipulation of nuclear-spin quantum states and can greatly facilitate the exploitation of the excellent properties of noble-gas nuclear spins in quantum science and technology.

Based on spin-exchange collisions, we develop a feasible method of generating and manipulating nuclear spin squeezing with an ultra-long lifetime and huge atomic number, by the coherent squeezing transfer between alkali-metal and noble-gas spin ensembles.
This method shows considerable advantage in preparing the spin squeezing of the macroscopic atomic ensemble with a huge atomic number, and even the nuclear spin-squeezed state containing $10^{20}$ atoms or more can be efficiently generated with preexisting techniques.  
By controlling the on/off switch of the spin-exchange interaction via the magnetic field, the spin squeezing can be stored in collective nuclear spins, whose lifetime can reach several days at or above room temperature thanks to the isolation from environment.
Moreover, based on the coherent transfer of the SSS between alkali-metal and noble-gas ensembles, the squeezing degree of nuclear spins can be obtained through alkali-metal spins, which provides a promising approach to quantitatively measure nuclear spin squeezing.

The paper is organized as follows. 
In Sec.~\ref{Sec-2}, we introduce the theoretical model and method for generating nuclear spin squeezing.
In Sec.~\ref{Sec-3}, we introduce the manipulation strategies for the generation, storage and readout of nuclear spin squeezing.
In Sec.~\ref{Sec-4}, we analyze the effects of dissipation, external-field manipulation precision and other relevant factors, and discuss the experimental feasibility of the proposed scheme.
The last section is a summary.

\section{Generation of nuclear spin squeezing}
\label{Sec-2}

\begin{figure}
	\includegraphics[draft=false, width=0.8\columnwidth]{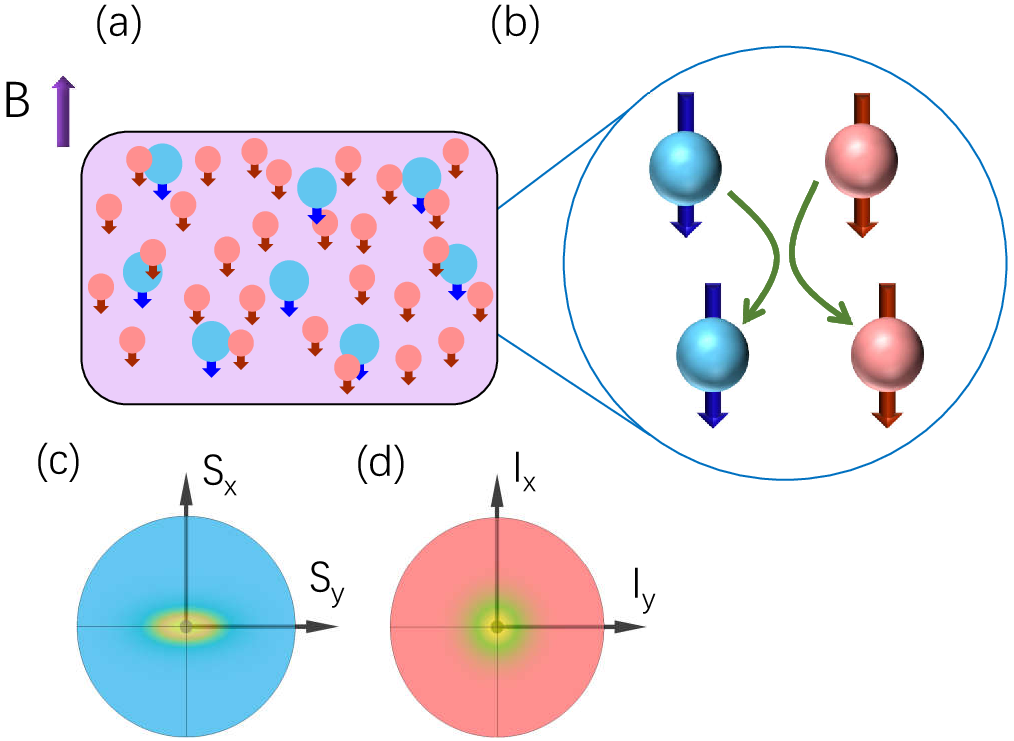}
	\caption{(Color online) (a) Schematic diagram of the alkali-metal (blue) and noble-gas (red) mixture with the spin-exchange collisions between the alkali-metal electronic spins and the noble-gas nuclear spins as shown in (b). The collective alkali and noble-gas spins are prepared initially to a SSS and CSS as shown in (c) and (d), respectively, with the collective spins along the same direction.
	\label{fig1}}
\end{figure}

\subsection{Squeezing transfer based on spin-exchange interaction}
\label{Sec-2a}

As shown in Fig.~\ref{fig1}, the mixed gases consisting of alkali-metal and noble-gas atoms are considered.
The coherent spin-exchange interaction between polarized alkali-metal spins and noble-gas nuclear spins can be expressed as an effective Hamiltonian~\cite{Katz2020_Long-Lived}
\begin{eqnarray}
{H_{se}} = \eta \mathbf{S} \cdot \mathbf{I}.       \label{eq-1}
\end{eqnarray}
without loss of generality~\cite{Katz2020_Long-Lived, Shaham2022_Strong, Katz2022_Quantum, Katz2022_Optical}. Here $\mathbf{S}=(S_x, S_y, S_z)$ and $\mathbf{I}=(I_x, I_y, I_z)$ denote the collective spin operators of $n_s$ alkali atoms and $n_i$ noble-gas atoms with spin-$1/2$, respectively, and $\eta$ denotes the spin-exchange rate.
The alkali and noble-gas spins are initially prepared to a SSS and coherent spin state (CSS), respectively, polarized against the direction of quantization axis ($\emph{z}$-axis), where the well-established QND method for the alkali spin squeezing~\cite{Bao2020_Spin, Kong2020_Measurement-induced, Kuzmich2000_Generation, Kuzmich1999_Quantum} can be employed.

\begin{figure}
	\includegraphics[draft=false, width=1\columnwidth]{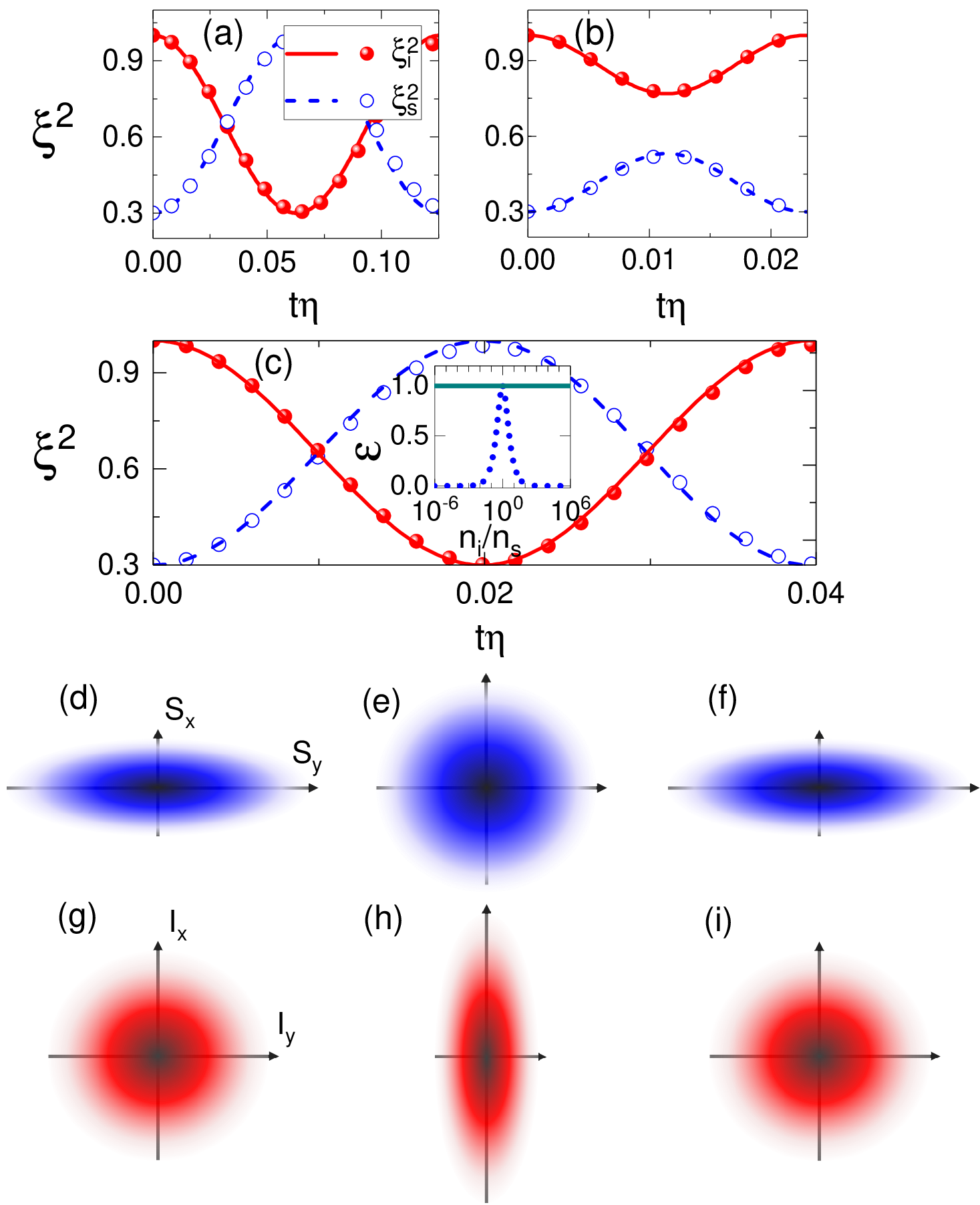}
	\caption{(Color online) Evolution of the spin-squeezing parameter under the Hamiltonian in Eq.~(\ref{eq-1}) for $n_s=50$, $n_i=50$ in (a), for $n_s=50$, $n_i=500$ in (b),  and under the Hamiltonian in Eq.~(\ref{eq-5}) with the magnetic field along $\emph{z}$-axis satisfying the resonance condition $(\gamma_s-\gamma_i)B=\frac{\eta }{2}(n_i - n_s)$ for $n_s=50$, $n_i=500$ in (c). Red (blue) curves and solid (open) circles denote the analytical and numerical results of the squeezing parameter of the noble-gas (alkali-metal) spins. The inset in (c) shows the transfer efficiency of spin squeezing defined as $\varepsilon=log_{10}\xi_{i,op}^2/log_{10}\xi_{s,o}^2$ without and with a magnetic field $B_r$ denoted by the blue dotted curve and green solid curve, respectively, as a function of $n_i/n_s$. Evolution of the quantum states of alkali and noble-gas spins represented by the Husimi-Q function at $t=0$, $\tau_{op}$ and $2\tau_{op}$ in (d)-(f) and (g)-(i), respectively, with the global rotation $R(t) = 0$.
	\label{fig2}}
\end{figure}

In Fig.~\ref{fig2}, we exhibit the evolution of squeezing parameters of the alkali and noble-gas spins. When $n_s=n_i$, as shown in Fig.~\ref{fig2}(a), the spin-squeezing parameters of the alkali and noble-gas spins oscillate sinusoidally, respectively, with the same amplitude. At one-half of a period, the spin-squeezing parameters of the two spin ensembles are completely exchanged, which demonstrates that it is feasible to prepare noble-gas nuclear spin squeezing by the squeezing transfer method.
However, when $n_s\ne n_i$, the squeezing transfer is hindered or even stopped, as shown in Fig.~\ref{fig2}. 
Therefore, the preparation of nuclear spin squeezing is limited when the atomic numbers of alkali and noble-gas ensembles are different.

To facilitate insight into the evolution of spin squeezing, we can perform the Holstein-Primakoff transformation as follows~\cite{Holstein1940_Field}
\begin{eqnarray}
&&{S_x} = \sqrt {{n_s}} {X_a},{S_y} = \sqrt {{n_s}} {P_a},{S_z} = {a^\dag }a - \frac{n_s}{2}, \nonumber   \\
&&{I_x} = \sqrt {{n_i}} {X_b},{I_y} = \sqrt {{n_i}} {P_b}, {I_z} = {b^\dag }b - \frac{n_i}{2},        \label{eq-2}
\end{eqnarray}
where ${X_k} = \frac{1}{2}(k + {k^\dag }),{P_k} =  - \frac{i}{2}(k - {k^\dag })$ are the canonical position and momentum operators with $k=a,b$ obeying the commutation relation $[k,{k^\dag }] = 1$. 
Therefore, the Hamiltonian~(\ref{eq-1}) becomes 
\begin{eqnarray}
{H'_{eff}} =  - \frac{1}{2}\eta n_i {a^\dag }a - \frac{1}{2}\eta n_s{b^\dag }b + \frac{J}{2} (a{b^\dag } + b{a^\dag }) ,        \label{eq-3}
\end{eqnarray}
with collective exchange rate $J =  \eta \sqrt {n_s n_i}$. Eq.~(\ref{eq-3}) reveals that the $\emph{x}$ and $\emph{y}$ components of the coherent spin-exchange interaction of alkali and noble-gas spins can be treated as a beam-splitter interaction, and that the $\emph{z}$ component leads to an effective detuning between the two modes $a$ and $b$. 

By solving the motion equations and applying the definition of the squeezing degree in Ref.~\cite{Kitagawa1993_Squeezed} (See Appendix~\ref{App-1}), the squeezing parameters of alkali and noble-gas spins are derived, respectively, as 
\begin{eqnarray}
&&\xi_s^2(t) = \xi_{s,0}^2 + \frac{{2r(1 - \xi_{s,0}^2)}}{{{{(1 + r)}^2}}}(1 - \cos (\frac{{{n_s} + {n_i}}}{2}\eta t)),    \nonumber   \\
&&\xi_i^2(t) = 1 - \frac{{2r(1 - \xi_{s,0}^2)}}{{{{(1 + r)}^2}}}(1 - \cos (\frac{{{n_s} + {n_i}}}{2}\eta t)).  \label{eq-4}
\end{eqnarray}
Here $\xi_{s,0}^2$ denotes the initial squeezing parameter of alkali spins and $r = n_i/n_s$ denotes the ratio of the two atomic numbers. 
Eq.~(\ref{eq-4}) reveals that when the evolution time satisfies $t = 2\pi/(\eta{n_s} + \eta {n_i})$, the squeezing parameter of noble-gas spins reaches the minimum $((r-1)^2 + 4r\xi_{s,0}^2)/(1 + r)^2\geq \xi_{s,0}^2 $. 
We see that when $n_s=n_i$, the effective beam-splitter interaction Hamiltonian~(\ref{eq-3}) is resonant. Therefore, a complete squeezing exchange ensures that the noble-gas spins evolve to a SSS with the same squeezing parameter $\xi_{s,0}^2$ as the initial alkali spins. 
However, when $n_s\ne n_i$, this beam-splitter interaction is nonresonant, resulting in the squeezing parameter of the noble-gas spins not reaching $\xi_{s,0}^2$.
Further, when ${n_i}/{n_s} \gg 1$ or ${n_i}/{n_s} \ll 1$, the beam-splitter interaction will be in large detuning regime, and thus the squeezing transfer between the two spin ensembles is forbidden.

\subsection{Efficient generation of long-lived nuclear spin squeezing}
\label{Sec-2b}

To recover the squeezing transfer efficiency when atomic numbers are different, we can apply a magnetic field $\mathbf{B}=B \hat{z}$ in the atomic ensembles along $\emph{z}$-axis.
Accordingly, the total Hamiltonian of the system becomes 
\begin{equation}
{H_{eff}} = ({\gamma_s B} - \frac{\eta n_i}{2}){a^\dag }a + ({\gamma_i B} - \frac{\eta n_s}{2}){b^\dag }b + \frac{J}{2} (a{b^\dag } + b{a^\dag }),   
\label{eq-5}
\end{equation}
where $\gamma_s$ and $\gamma_i$ are the gyromagnetic ratios of alkali and noble-gas spins, respectively.
Eq.~(\ref{eq-5}) demonstrates that when the magnetic field $\mathbf{B}$ has a specific strength 
\begin{equation}
B_r=\frac{\eta}{2}\frac{n_i - n_s}{\gamma_s -\gamma_i}, 
\label{eq-5b}
\end{equation}
the effective beam-splitter interaction between two ensembles will remain resonant regardless of different atomic numbers. 
According to the derivations in Appendix~\ref{App-1}, the spin-squeezing parameters of alkali and noble-gas spins are, respectively, obtained as
\begin{eqnarray}
\xi_s^2(t) &=\frac{1}{2} \left(1 + \xi_{s,0}^{2} - (1 - \xi_{s,0}^{2}) \cos(Jt)\right)  \nonumber  \\
\xi_i^{2}(t) &= \frac{1}{2} \left(1 + \xi_{s,0}^{2} + (1 - \xi_{s,0}^{2}) \cos(Jt)\right)    \label{eq-6}
\end{eqnarray} 
Eq.~(\ref{eq-6}) reveals that under the above scheme, the noble-gas spins can evolve to a SSS with the same squeezing parameter $\xi_{s,0}^{2}$ as the initial alkali-metal spin squeezing, even though the numbers of the two spin ensembles are significantly different, as visualized in Fig.~\ref{fig2}(c). The corresponding optimal evolution time satisfies $\tau_{op} = \pi {J^{-1}}$.  
Thereby, by controlling the magnetic field $\mathbf{B}$, efficient squeezing transfer from alkali spins to noble-gas spins is achieved regardless of different atomic numbers.
Moreover, we demonstrate that this method is valid over a large squeezing range due to huge atomic numbers of spin ensembles (See Appendix~\ref{App-2}).
We show the Husimi-Q functions~\cite{Kitagawa1993_Squeezed} of the alkali and noble-gas spins in Figs.~\ref{fig2}(d)-(i), visualizing the evolution of quantum states.
Starting from the initial states, the alkali and noble-gas spins exchange the spin states when $t=\tau_{op}$, thereby the noble-gas spins evolve to the SSS with squeezing parameter $\xi_{s,0}^2$. When $t=2\tau_{op}$, these two spin ensembles return to the initial states by exchanging states again.  
During the evolution, the squeezing directions of the alkali and noble-gas spins remain perpendicular, due to the phase induced by the beam-splitter interaction~(\ref{eq-5}) between two spin ensembles. 
Besides, the squeezing directions undergo a global rotation $R(t) = \delta_r t$, where $\delta _r = \gamma_s B_r - \eta {n_i}/2 = \gamma_i B_r - \eta {n_s}/2$ is the global frequency shift of the two ensembles. 
When ${n_i}/{n_s} = {\gamma_s}/{\gamma_i}$, the rotation $R(t) = 0$, i.e., the spin squeezings of the alkali and noble-gas atoms are always along x- and y-axes, respectively, as shown in Figs.~\ref{fig2}(d)-(i).
Unless otherwise specified, all subsequent discussions are performed in the rotating frame at frequency $\delta _r$ or under the condition $\delta_r=0$, whereby the trivial global rotation is omitted.

\section{Coherent manipulation}
\label{Sec-3}

\begin{figure}
	\includegraphics[draft=false, width=1\columnwidth]{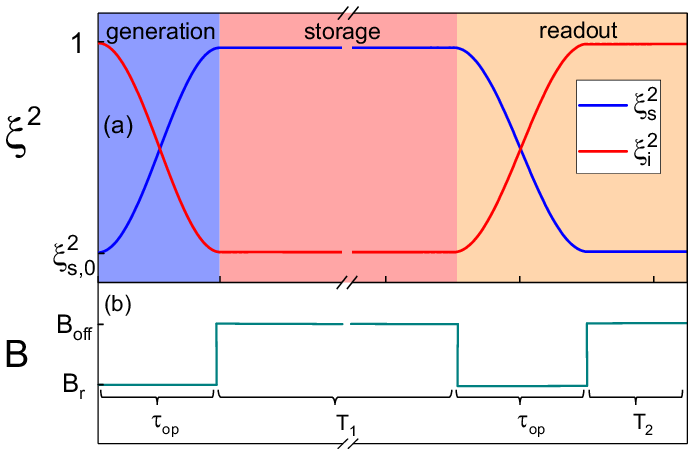}
	\caption{(Color online) (a) Evolution of the spin-squeezing parameters manipulated by the magnetic field sequence shown in (b). Generating nuclear spin squeezing (blue region): Setting $B = B_{r}$, the squeezing transfer is turned on. After the duration $\tau_{op}$, the noble-gas nuclear spins will evolve to a SSS with the same squeezing parameter as the initial alkali-metal spins. 
	Storing nuclear spin squeezing (red region): Setting $B = B_{o\! f\! f}$ at $t=\tau_{op}$, the squeezing transfer between alkali and noble-gas spins is turned off. Thus, the spin squeezing resides in the long-lifetime noble-gas nuclear spins. 
	Measuring nuclear spin squeezing (orange region): Setting $B = B_{r}$, the alkali spins in the probe state (CSS) couple with the noble-gas nuclear spins which is in the SSS with a squeezing parameter $\xi_{i,m}^{2}$ to be measured. After a duration $\tau_{op}$, the alkali spins will evolve to the SSS with the squeezing parameter $\xi_{i,m}^{2}$. Next, setting $B = B_{o\! f\! f}$, the alkali spin squeezing will be maintained, and thus the squeezing parameter $\xi_{i,m}^{2}$ of nuclear spins can be readout by measuring the squeezing degree of alkali spins. 
\label{fig3}}
\end{figure}
It is worth noting that the coherent squeezing transfer process between alkali and noble-gas spins can be easily turned on and off via the magnetic field $\mathbf{B}$. When $B = B_{r}$, the spin squeezing can be completely transferred between alkali and noble-gas spins.
On the contrary, when $B = B_{o\! f\! f}$  with $\left| ({\gamma_s} - {\gamma_i}){B_{o\! f\! f}} - \eta({n_i} - {n_s})/2 \right| \gg J/2 $, or the atomic numbers satisfy ${n_i}/{n_s} \gg 1$ or ${n_i}/{n_s} \ll 1$, the squeezing transfer is turned off. 
The QND method for obtaining alkali-metal spin squeezing is naturally compatible with the preparation process of nuclear SSS (See Appendix~\ref{App-3}), and the complete process can be described by the Hamiltonian $H_{Q} = {\gamma_s B(t)} {S_z} + {\gamma_i B(t)} {I_z} + \frac{k(t)}{t_p} J_x S_x + \eta \mathbf{S} \cdot \mathbf{I}$.
The third term describes the QND (Faraday) interaction between the probe light pulse with photon number $n_{ph}$ and alkali-metal spins, with the interaction strength $k$, and the Stokes operator $J_x$ and duration $t_p$ of the light. 
Both spin ensembles start with CSSs polarized against $\emph{z}$-axis. For atomic vapor typically satisfying ${n_i}/{n_s} \gg 1$, the interaction between the alkali and noble-gas spins is turned off at $B=0$, and thus, only the QND interaction term of $H_{Q}$ comes into play. Therefore, according to the QND method for the preparation of spin squeezing~\cite{Bao2020_Spin, Kong2020_Measurement-induced, Kuzmich2000_Generation, Kuzmich1999_Quantum}, the alkali-metal spin squeezing, whose squeezing parameter satisfies $\xi_{s, q}^2 = 1/(1+\kappa^2)$ with $\kappa^2= k^2 n_s n_{ph}$, can be obtained via the measurement of $J_x$. Next, for the light pulse leaves the atomic gases, $H_{Q}$ will degenerate to the resonant Hamiltonian in Eq.~(\ref{eq-5}) by setting $B = B_{r}$. After a duration $\tau_{op}$, the noble-gas spin squeezing with squeezing parameter $\xi_{s, q}^2$ will be obtained.

As shown in Fig.~\ref{fig3}, when the preparation of noble-gas spin squeezing is finished, the squeezing transfer between two spin ensembles can be turned off via the magnetic field $\mathbf{B}$. 
Therefore, the spin squeezing is stored in the noble-gas nuclear spins and can be maintained for several days thanks to the isolation from environment.
Moreover, alkali spins can be used as a probe to measure the squeezing degree of the noble-gas nuclear spins, where the measurement is an inverse process of generating the nuclear spin squeezing. Specifically, the alkali spins are needed to be initially prepared to the probe state, i.e., the CSS along the same direction as the noble-gas spins. By setting $B = B_{r}$, the squeezing transfer between alkali and noble-gas spins is turned on, and after the time $\tau_{op}$, the squeezing parameter of the noble-gas spin will be mapped onto the alkali spins, as shown in Fig.~\ref{fig3}.
After turning off squeezing transfer, the specific squeezing parameter of noble-gas spins can be obtained by measuring the squeezing parameter of the alkali spins, which is measurable based on current experimental techniques~\cite{Hald1999_Spin, Kuzmich1999_Quantum, Kuzmich2000_Generation, Sewell2012_Magnetic, Koschorreck2010_Sub-Projection-Noise, Hosten2016_Measurement, Bao2020_Spin}. 
Therefore, by adjusting the strength of the magnetic field $\mathbf{B}$, all the preparation of the initial alkali spin squeezing and the preparation, storage, and readout of the noble-gas nuclear spin squeezing can be independently carried out.

\section{Experimental feasibility}
\label{Sec-4}

\begin{figure}
	\includegraphics[draft=false, width=0.95\columnwidth]{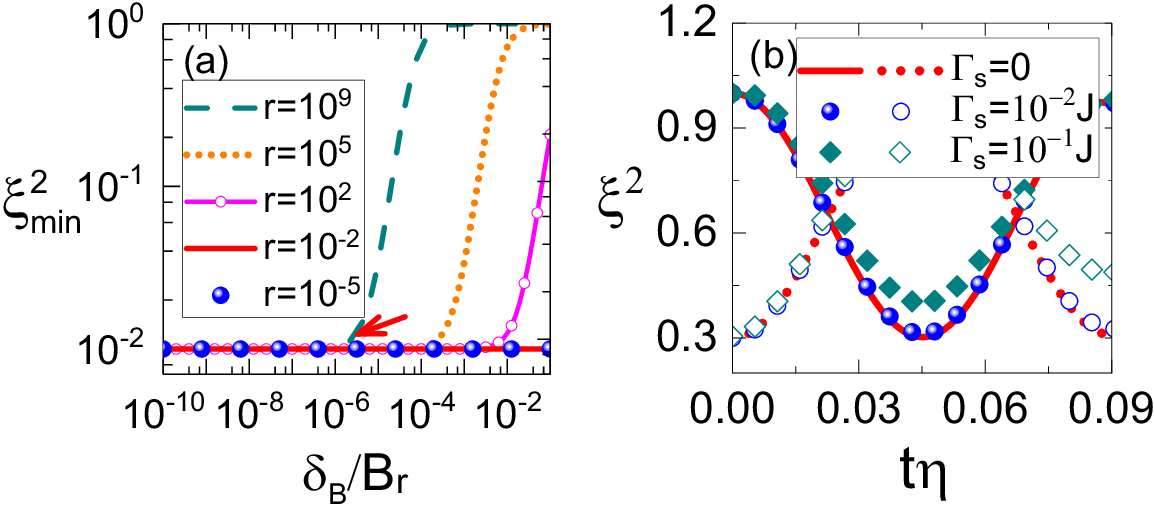}
	\caption{(Color online)  (a) The optimal squeezing parameter of noble-gas nuclear spins as a function of the deviation $\delta_B$ of the magnetic field from $B_r$ for different ratios of atomic numbers $r=n_i/n_s$, with the initial squeezing parameter of alkali spins $\xi_{s,0}^2=10^{-2}$. 
 According to the curve for $n_i/n_s=10^9$ and Eq.~(\ref{eq-5b}), if $J\approx 80 Hz$, a magnetic field accuracy of about $1 nT$ is enough for efficient transfer of spin squeezing from alkali spins to noble-gas spins, as marked by the red arrow. 
 (b) Evolution of the squeezing parameters of alkali and noble-gas spins for different relaxation rate $\Gamma_s$ of alkali spins with $n_s=50$, $n_i=100$. The relaxation rate of noble-gas spins is several orders of magnitude smaller than that of alkali spins, and thus is neglected.
\label{fig4}}
\end{figure}

To achieve complete squeezing transfer, the magnetic field strength needs to satisfy the resonance condition in Eq.~(\ref{eq-5b}). Considering the limited accuracy of a magnetic field in experiments, we plot in Fig.~\ref{fig4}(a) the optimal squeezing parameter of noble-gas spins as a function of the deviation $\delta_B$ of the magnetic field from $B_r$ for different ratios of atomic numbers. 
One can deduce that for a vapor cell with $J\approx 80 Hz$~\cite{Shaham2022_Strong} and a magnetic field deviation about $nT$, a high transfer efficiency of spin squeezing between alkali and noble-gas spins can be ensured even if the atomic numbers of the two spin systems differ by many orders of magnitude, e.g., $n_i/n_s=10^9$, as marked in Fig.~\ref{fig4}(a)  (red arrow). 
Significantly, this proposal shows a considerable advantage in preparing the spin squeezing of the macroscopic atomic ensemble with a huge atomic number.
Refs.~\cite{Bao2020_Spin, Kong2020_Measurement-induced} report spin squeezing containing $10^{11} \sim 10^{13}$ alkali-metal atoms in a hot vapor cell based on QND method. Accordingly, the squeezing of $10^{20}$ or even more noble-gas nuclear spins is feasible based on preexisting techniques. 
In Fig.~\ref{fig4}(b), we also exhibit the evolution of the squeezing parameter when the relaxation processes are involved. It reveals that the strong coupling condition of the coherent spin-exchange interaction, i.e., $J \gg {\Gamma _s}$, is required for the high transfer efficiency and measurement accuracy of the spin squeezing. 
According to Ref.~\cite{Shaham2022_Strong}, this condition is realizable in a hot ${}^3$He-K mixture enclosed in a vapor cell, with the coupling-to-decay ratio $J/\Gamma_s\approx 10$, and a larger coupling-to-decay ratio is also expected by optimizing the experimental parameters. 
The preceding discussion neglected the effects of imperfect polarization. For a comprehensive understanding, we investigate the effects of imperfect polarization in Appendix~\ref{App-4}.
It is indicated that the squeezing transfer process can be well maintained under imperfect polarization when the coherent spin-exchange interaction operates in the strong-coupling regime. 
Furthermore, near-unity polarization of alkali-metal spins has been reported in Ref.~\cite{Dideriksen2021_Room-temperature}, and 85\% polarization of $^3$He has been achieved in Ref.~\cite{Chen2014_On}.

\section{Conclusion}
\label{Sec-5}

In this work we propose a feasible method of generating and manipulating the nuclear spin squeezing with ultra-long lifetime and huge atomic number. 
By the coherent squeezing transfer based on spin-exchange collisions and the manipulation of a magnetic field, we find that the nuclear SSS can be efficiently obtained.
This proposal has considerable advantage in preparing the spin squeezing of the macroscopic atomic ensemble with a huge atomic number, and the nuclear spin-squeezed state containing $10^{20}$ or even more atoms can be generated based on preexisting techniques.
The spin squeezing can be stored in the nuclear spin ensemble with a days-long lifetime and the squeezing degree of the nuclear spins can be measured via an inverse squeezing transfer process.
This proposal can be implemented in hot atomic ensembles contained in vapor cells, which is widely used in quantum metrology, fundamental physics, and medicine, due to various advantages, including long life, miniaturization potential, and high adaptability to various environments. 
Moreover, by simple manipulation of the magnetic field, all the preparation of alkali-metal SSS, and the preparation, storage, and measurement of noble-gas nuclear SSS can be independently carried out, which makes the proposal easy to perform and apply.
The obtaining and manipulation of macroscopic nuclear-spin quantum states with ultra-long lifetimes enable quantum technologies with high stability and efficiency, and will significantly facilitate the research in quantum metrology and quantum information science.

\begin{acknowledgments}
This work is supported by the National Key R\&D Program of China (Grant No. 2023YFA1407600), the National Natural Science Foundation of China (NSFC) (Grants No. 92576204, No. 12275145, No. 92050110, No. 91736106, No. 11674390, and No. 91836302), and Beijing Key Laboratory of Quantum Artificial Intelligence.
\end{acknowledgments}

\appendix

\section{Squeezing parameters of alkali-metal and noble-gas spins}
\label{App-1}

Under the Holstein-Primakoff approximation, the spin components in the $x$- and $y$-directions can be transformed into the canonical position and momentum operators, respectively, 
\begin{align}
	S_x &= \sqrt {{n_s}} {X_a},  \quad 
	S_y = \sqrt {{n_s}} {P_a},   \nonumber   \\
	I_x &= \sqrt {{n_i}} {X_b},   \hspace{4.5mm}   
	I_y = \sqrt {{n_i}} {P_b},    
	\label{eq-S7}
\end{align}
where $[X_a,P_a]=[X_b,P_b] = \frac{i}{2}$. The spin components in the $z$-direction can be approximated by the mean values $S_z \approx \langle S_{z} \rangle$ and $I_z \approx \langle I_{z} \rangle$.  
According to the Hamiltonian (\ref{eq-5}), the explicit form of the Heisenberg equations for the mean value of the canonical position and momentum operators can be expressed as 

\begin{align}
	\partial_t \langle X_a \rangle &= -\Delta_s \langle P_a \rangle - \frac{1}{2} J \langle P_b \rangle, \nonumber   \\
	\partial_t \langle P_a \rangle &= \Delta_s \langle X_a \rangle + \frac{1}{2} J \langle X_b \rangle, \nonumber   \\
	\partial_t \langle X_b \rangle &= -\Delta_i \langle P_b \rangle - \frac{1}{2} J \langle P_a \rangle, \nonumber   \\
	\partial_t \langle P_b \rangle &= \Delta_i \langle X_b \rangle + \frac{1}{2} J \langle X_a \rangle,
	\label{eq-S8}
\end{align}
with $\Delta_s = \gamma_s B - n_i \eta /2$, $\Delta_i = \gamma_i B - n_s \eta/2 $.

Considering that the initial mean spins of alkali and noble-gas spins are opposite to the direction of the $\emph{z}$-axis, one can deduce according to Eq.~(\ref{eq-S8}) that the direction of the mean spins remains constant. We introduce the spin components normal to the mean spins as 
\begin{align}
	F_{\perp} = F_{x}\cos(\alpha) + F_{y}\sin(\alpha)   ,
	\label{eq-S9}
\end{align}
with $F = S, I$. Therefore, the variance of $F_{\perp}$ reads 
\begin{align}
	(\Delta {F}_{\perp})^{2} = \frac{1}{2} (\mathcal{C}_{f} + \mathcal{A}_{f}\cos(2\alpha) + \mathcal{B}_{f}\sin(2\alpha)),
	\label{eq-S10}
\end{align}
where the coefficients $\mathcal{A}_{f}, \mathcal{B}_{f}$ and $\mathcal{C}_{f}$ are defined as 
\begin{align}
	\mathcal{A}_{f} &= \langle {F}_{x}^{2} - {F}_{y}^{2} \rangle ,  \nonumber   \\
	\mathcal{B}_{f} &= \langle {F}_{x}{F}_{y} + {F}_{y}{F}_{x}  \rangle ,    \nonumber   \\
	\mathcal{C}_{f} &= \langle {F}_{x}^{2} + {F}_{y}^{2} \rangle  .
	\label{eq-S11}
\end{align}

By minimizing Eq.~(\ref{eq-S10}) with respect to $\alpha$, the optimal squeezed angle $\alpha_{op}$ can be obtained, yielding
\begin{align}
	\tan(2\alpha_{op}) = \frac{\mathcal{B}}{\mathcal{A}}, 
	\label{eq-S12}
\end{align}
therefore, $\cos(2\alpha) = \pm \mathcal{A}/\sqrt{\mathcal{A}^2 + \mathcal{B}^2}$,  
	$\sin(2\alpha) = \pm \mathcal{B}/\sqrt{\mathcal{A}^2 + \mathcal{B}^2} $.
Substituting these results into Eq.~(\ref{eq-S10}), we obtain the  the minimum variance of $F_{\perp}$
\begin{align}
	(\Delta {F}_{\perp})^{2}_{min} &= \frac{1}{2} (\mathcal{C}_{f} - \sqrt{\mathcal{A}_{f}^{2} + \mathcal{B}_{f}^{2}})
	\label{eq-S13}
\end{align}
According to Ref.~\cite{Kitagawa1993_Squeezed}, the degree of spin squeezing can be quantified by the squeezing parameter
\begin{align}
	\xi^2_{f}=\frac{2(\Delta {F}_{\perp})^{2}_{min}}{\langle F \rangle} = \frac{2}{n_f} (\mathcal{C}_{f} + \sqrt{\mathcal{A}_{f}^{2} + \mathcal{B}_{f}^{2}}).
	\label{eq-S14}
\end{align}

To obtain the explicit expressions of the squeezing parameters of alkali and noble-gas spins, we next focus on the time evolution of the system. 
Under the Holstein-Primakoff approximation, the coefficients $\mathcal{A}_s, \mathcal{B}_s, \mathcal{C}_s $ and $\mathcal{A}_i, \mathcal{B}_i, \mathcal{C}_i $ in Eq.~(\ref{eq-S11}) can be expressed as 
\begin{align}
	\mathcal{A}_s &= n_s \langle X_a^2 - P_a^2 \rangle,  \nonumber   \\
	\mathcal{B}_s &= n_s \langle X_a P_a + P_a X_a \rangle,   \nonumber   \\
	\mathcal{C}_s &= n_s \langle X_a^2 + P_a^2 \rangle    \nonumber   \\
	\mathcal{A}_i &= n_i \langle X_b^2 - P_b^2 \rangle,   \nonumber   \\
	\mathcal{B}_i &= n_i \langle X_b P_b + P_b X_b \rangle,   \nonumber   \\
	\mathcal{C}_i &= n_i \langle X_b^2 + P_b^2 \rangle .
	\label{eq-S15}
\end{align}
According to the Hamiltonian~(\ref{eq-5}), the motion equations for the mean values of the above operators are given by  
\begin{align}
	\partial_t \langle X_a^2 \rangle =& -\frac{i}{2} \Delta_s - 2\Delta_s \langle X_a P_a \rangle - J \langle X_a P_b \rangle, \nonumber   \\
	\partial_t \langle X_a P_a \rangle =& -\Delta_s \langle P_a^2 \rangle + \Delta_s \langle X_a^2 \rangle - \frac{1}{2} J \langle P_a P_b \rangle + \frac{1}{2} J \langle X_a X_b \rangle, \nonumber   \\
	\partial_t \langle X_a X_b \rangle =& -\Delta_i \langle X_a P_b \rangle - \Delta_s \langle P_a X_b \rangle - \frac{i}{4} J - \frac{1}{2} J \langle X_b P_b \rangle    \nonumber   \\
	    &  - \frac{1}{2} J \langle X_a P_a \rangle, \nonumber   \\
	\partial_t \langle X_a P_b \rangle =& \Delta_i \langle X_a X_b \rangle - \Delta_s \langle P_a P_b \rangle - \frac{1}{2} J \langle P_b^2 \rangle + \frac{1}{2} J \langle X_a^2 \rangle, \nonumber   \\
	\partial_t \langle P_a^2 \rangle =& \frac{i}{2} \Delta_s + 2\Delta_s \langle X_a P_a \rangle + J \langle P_a X_b \rangle, \nonumber   \\
	\partial_t \langle P_a X_b \rangle =& -\Delta_i \langle P_a P_b \rangle + \Delta_s \langle X_a X_b \rangle - \frac{1}{2} J \langle P_a^2 \rangle            \nonumber   \\
	\partial_t \langle P_a P_b \rangle =&   \Delta_i \langle P_a X_b \rangle + \Delta_s \langle X_a P_b \rangle + \frac{i}{4} J + \frac{1}{2} J \langle X_b P_b \rangle         \nonumber   \\
	      &+ \frac{1}{2} J \langle X_a P_a \rangle,            \nonumber   \\
	\partial_t \langle X_b^2 \rangle =& -\frac{i}{2} \Delta_i - 2\Delta_i \langle X_b P_b \rangle - J \langle P_a X_b \rangle, \nonumber   \\
	\partial_t \langle X_b P_b \rangle =& -\Delta_i \langle P_b^2 \rangle + \Delta_i \langle X_b^2 \rangle - \frac{1}{2} J \langle P_a P_b \rangle + \frac{1}{2} J \langle X_a X_b \rangle, \nonumber   \\
	\partial_t \langle P_b^2 \rangle =& \frac{i}{2} \Delta_i + 2\Delta_i \langle X_b P_b \rangle + J \langle X_a P_b \rangle. 
	\label{eq-S16}
\end{align}

Firstly, we consider the case without an external magnetic field, $B=0$. By solving Eq.~(\ref{eq-S16}), one can obtain the explicit expressions of the coefficients 
\begin{widetext}
\begin{align}
	\mathcal{A}_s &= \frac{n_s ((\xi_{s, 0}^2)^2 - 1) (n_s^2 + 2 n_s n_i \cos(\frac{1}{2} (n_s + n_i) \eta t) + n_i^2 \cos((n_s + n_i) \eta t))}{4 (n_s + n_i)^2 \xi_{s, 0}^2},   \nonumber   \\
	\mathcal{B}_s &= -\frac{n_s n_i ((\xi_{s, 0}^2)^2-1) (2 n_s \sin(\frac{1}{2} (n_s + n_i) \eta t) + n_i \sin((n_s + n_i) \eta t))}{4 (n_s + n_i)^2 \xi_{s, 0}^2},   \nonumber   \\
	\mathcal{C}_s &= \frac{n_s (4 n_s n_i \xi_{s, 0}^2 + (n_s^2 + n_i^2) (1 + (\xi_{s, 0}^2)^2) + 2 n_s n_i (\xi_{s, 0}^2 - 1)^2 \cos(\frac{1}{2} (n_s + n_i) \eta t))}{4 (n_s + n_i)^2 \xi_{s, 0}^2}  \nonumber   \\
	\mathcal{A}_i &= -\frac{n_s n_i^2 ((\xi_{s, 0}^2)^2 - 1) \cos(\frac{1}{2} (n_s + n_i) \eta t) \sin(\frac{1}{4} (n_s + n_i) \eta t)^2}{(n_s + n_i)^2 \xi_{s, 0}^2},    \nonumber   \\
	\mathcal{B}_i &= \frac{n_s n_i^2 ((\xi_{s, 0}^2)^2-1) \sin(\frac{1}{4} (n_s + n_i) \eta t)^2 \sin(\frac{1}{2} (n_s + n_i) \eta t)}{(n_s + n_i)^2 \xi_{s, 0}^2},    \nonumber   \\
	\mathcal{C}_i &= \frac{n_i ((n_i + n_s \xi_{s, 0}^2) (n_s + n_i \xi_{s, 0}^2) - n_s n_i (\xi_{s, 0}^2 - 1)^2 \cos(\frac{1}{2} (n_s + n_i) \eta t))}{2 (n_s + n_i)^2 \xi_{s, 0}^2},
	\label{eq-S17}
\end{align}
\end{widetext}
with $\xi_{s,0}^2$ the initial squeezing parameter of alkali spins $\mathbf{S}$.
Substituting Eq.~(\ref{eq-S17}) into Eq.~(\ref{eq-S14}), we obtain the squeezing parameters of alkali and noble-gas spins, respectively, as
\begin{align}
	\xi_s^2(t) &= \xi_{s,0}^2 + \frac{{2r(1 - \xi_{s,0}^2)}}{{{{(1 + r)}^2}}}(1 - \cos (\frac{{{
				n_s} + {n_i}}}{2}\eta t)),    \nonumber   \\
	\xi_i^2(t) &= 1 - \frac{{2r(1 - \xi_{s,0}^2)}}{{{{(1 + r)}^2}}}(1 - \cos (\frac{{{n_s} + {n_i}}}{2}\eta t)), 
	\label{eq-S18}
\end{align}
with $r = n_i/n_s$.

By applying a magnetic field $\mathbf{B}$ along the $\emph{z}$-axis satisfying $B=\eta (n_i-n_s)/(2\gamma_s-2\gamma_i)$, the effective detuning between the alkali and noble-gas spins can be eliminated. Besides, the trivial global frequency shift can also be eliminated by a transformation of the rotating frame. Therefore, we solve Eq.~(\ref{eq-S16}) and obtain the explicit expressions of the coefficients
\begin{align}
	\mathcal{A}_s &= \frac{n_s ((\xi_{s,0}^2)^2-1)  (1+\cos(J t))}{8 \xi_{s,0}^2},  \nonumber   \\
	\mathcal{B}_s &= 0,   \nonumber   \\
	\mathcal{C}_s &= \frac{n_s ((1+\xi_{s,0}^2)^2+(1-\xi_{s,0}^2)^2 \cos(J t))}{8 \xi_{s,0}^2}  \nonumber   \\
	\mathcal{A}_i &= \frac{n_i (1-(\xi_{s,0}^2)^2)  (1-\cos(J t))}{8\xi_{s,0}^2},   \nonumber   \\
	\mathcal{B}_i &= 0,   \nonumber   \\
	\mathcal{C}_i &= \frac{n_i ((1+\xi_{s,0}^2)^2-(1-\xi_{s,0}^2)^2 \cos(J t))}{8 \xi_{s,0}^2}.
	\label{eq-S19}
\end{align}
Substituting Eq.~(\ref{eq-S19}) into Eq.~(\ref{eq-S14}), we obtain the squeezing parameters of alkali and noble-gas spins, respectively, as 
\begin{align}
	\xi_s^2(t) &=\frac{1}{2} \left(1 + \xi_{s,0}^{2} - (1 - \xi_{s,0}^{2}) \cos(Jt)\right) \nonumber   \\
	\xi_i^{2}(t) &= \frac{1}{2} \left(1 + \xi_{s,0}^{2} + (1 - \xi_{s,0}^{2}) \cos(Jt)\right)    .     
	\label{eq-S20}
\end{align} 
Moreover, the optimal squeezed angles $\alpha_{s, op}$ and $\alpha_{i, op}$ of the alkali and noble-gas spins satisfy  
\begin{align}
	tan(2\alpha_{s,op}) = tan(2\alpha_{i,op}) =  0,
	\label{eq-S21}
\end{align} 
%which means that the spin squeezings of the alkali and noble-gas ensembles are in either the $x$- or $y$-direction. 
Thus, the optimal squeezed angles $\alpha_{s, op}$ and $\alpha_{i, op}$ are equal to either 0 or $\pi/2$.
To determine the specific directions of the spin squeezings, we derive the squeezing parameters $\xi_{f,x}^{2} = 4(\Delta F_{x})^{2}/n_{f}, \xi_{f,y}^{2} = 4(\Delta F_{y})^{2}/n_{f}$ of the spin components $F_x$ and $F_y$ 
by solving Eq.~(\ref{eq-S16}) and obtain
\begin{align}
	\xi_{s,x}^{2}(t) &= \frac{1}{2} \left(1 + \xi_{s,0}^{2} - (1 - \xi_{s,0}^{2}) \cos(Jt)\right) \nonumber   \\
	\xi_{s,y}^{2}(t) &= \frac{1 + \xi_{s,0}^{2} + (1 - \xi_{s,0}^{2}) \cos(Jt)}{2 \xi_{s,0}^{2}} \nonumber   \\
	\xi_{i,x}^{2}(t) &= \frac{1 + \xi_{s,0}^{2} - (1 - \xi_{s,0}^{2}) \cos(Jt)}{2 \xi_{s,0}^{2}} \nonumber   \\
	\xi_{i,y}^{2}(t) &= \frac{1}{2} \left(1 + \xi_{s,0}^{2} + (1 - \xi_{s,0}^{2}) \cos(Jt)\right)  .
	\label{eq-S22}
\end{align}
Comparing Eqs.~(\ref{eq-S20}) and (\ref{eq-S22}), one can see that the optimal spin squeezings of the alkali spins $\mathbf{S}$ and noble-gas spins $\mathbf{I}$ are consistent with the directions of the spin components $S_x$ and $I_y$, respectively.

\section{Constraint on squeezing degree}
\label{App-2}

\begin{figure}[htbp]
	\centering
	\includegraphics[width=5cm]{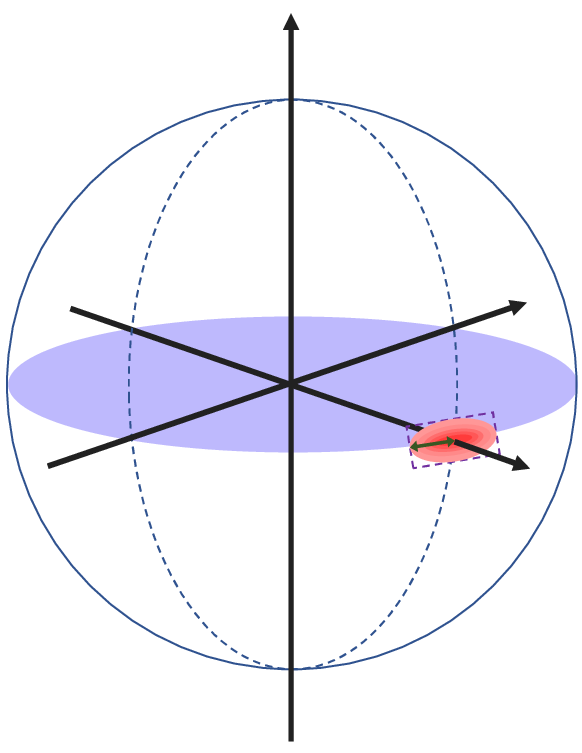}
	\caption{(Color online) Schematic diagram of the quasiprobability distribution of a SSS on the Bloch sphere. The green double arrows indicate the span of the spin fluctuation in the direction of the anti-squeezing.}
	\label{fig-S1}
\end{figure}

In the main text, we propose the method of generating nuclear spin squeezing by transferring coherently the SSS of alkali-metal spins. 
For the spin ensembles we consider, the direction and magnitude of the collective spin $\mathbf{F}$ ($\mathbf{F}= \mathbf{S}, \mathbf{I}$) remain approximately constant. Moreover, the excitation of the spin ensemble is much smaller than the total spin to ensure that the conditions for the Holstein-Primakoff approximation are satisfied. Geometrically, the above conditions require that the span of the distribution range of the spin component perpendicular to the polarization is much smaller than the radius of the Bloch sphere, so that the quasiprobability distribution on the Bloch sphere can be approximated as being in a plane. Therefore, the fluctuation in the direction of the anti-squeezing should satisfies  
\begin{eqnarray}
	\left\langle {\Delta F_{\max }^2} \right\rangle  \ll \left\langle {F_z^2} \right\rangle  \approx \frac{{n_f^2}}{4} ,
	\label{eq-S1}
\end{eqnarray}
or the following form
\begin{eqnarray}
	\xi _{f,\max }^2 \ll {n_f}  ,
	\label{eq-S2}
\end{eqnarray}
where $\xi _{f,\max }^2$ denotes the spin-squeezing parameter in the direction of the anti-squeezing, as shown in Fig.~\ref{fig-S1}, and $n_f$ denotes the number of atoms contained in the ensemble corresponding to F, with $f=s,i$. According to the Heisenberg uncertainty relation, the fluctuations in the directions of the optimal squeezing and anti-squeezing satisfy $\left\langle {\Delta F_{\min }^2} \right\rangle \left\langle {\Delta F_{\max }^2} \right\rangle  = \left\langle {F_z^2} \right\rangle /4$ for the minimum uncertainty state, that is, 
\begin{eqnarray}
	\xi _{f,\min }^2\xi _{f,\max }^2 = 1  .
	\label{eq-S3}
\end{eqnarray}
According to Eqs.~(\ref{eq-S2}) and (\ref{eq-S3}), we obtain the following formula
\begin{eqnarray}
	\xi _{f,\min}^2 \gg \frac{1}{{{n_f}}}.
	\label{eq-S4}
\end{eqnarray}

In our work, the optimal squeezing of the spin ensembles is determined by the initial spin squeezing of alkali-metal spins, i.e.,  $\xi _{f,\min }^2 = \xi _{s,0}^2$,  and thus, according to Eq.~(\ref{eq-S4}), we can obtain
\begin{eqnarray}
	\xi _{s,0}^2 \gg \frac{1}{{{n_f}}}.
	\label{eq-S5}
\end{eqnarray}
Since the two spin ensembles composed of alkali-metal and noble-gas atoms are present in our system, we can obtain the explicit constraint on the spin-squeezing parameters of both ensembles as
\begin{eqnarray}
	\xi _{s}^2, \xi _{i}^2  \gg \frac{1}{{{n_s}}},\frac{1}{{{n_i}}}  .
	\label{eq-S6}
\end{eqnarray}
One can derive that due to huge atomic numbers of spin ensembles, the proposal for efficient squeezing transfer in the main text is valid over a large range of squeezing degree.
Moreover, since the Holstein-Primakoff approximation is frequently employed in studies on spin squeezing~\cite{Hammerer2010_Quantum, Serafin2021_Nuclear, Katz2020_Long-Lived, Bao2020_Spin, Vasilakis2015_Generation}, the conclusion in Eq.~(\ref{eq-S4}) are broadly applicable and can facilitate further understanding of related work.

\section{Details on preparation and manipulation of nuclear spin squeezing}
\label{App-3}

\begin{figure*}[htbp]
	\centering
	\includegraphics[width=14cm]{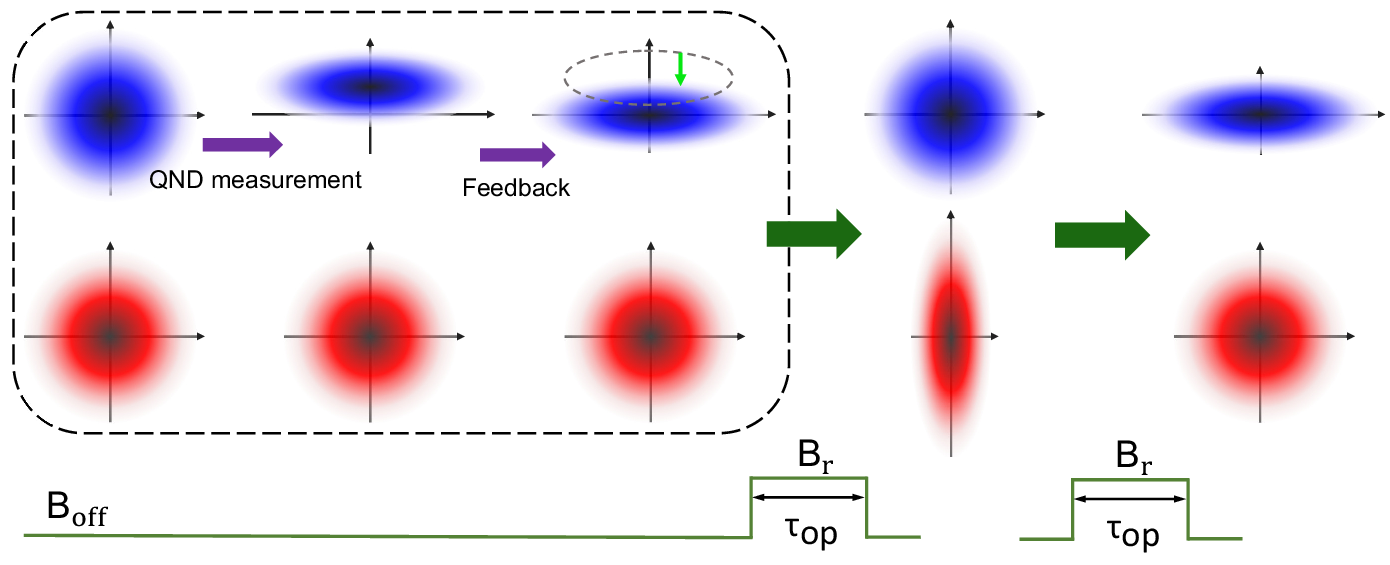}
	\caption{(Color online) 
		Sequence of preparation and measurement of nuclear spin squeezing starting from the pre-preparation of alkali-metal spin squeezing. Both the alkali-metal and noble-gas spins are initially prepared to the CSSs. As shown in the black box, the pre-preparation of the alkali-metal spin squeezing is based on the typical QND method. By detecting a light pulse through the atomic ensemble, the alkali-metal spins collapses to a conditional SSS with a random shift depending on the measurement result. The shift can be removed via feedback control, resulting in an unconditional SSS. During the above process, the noble-gas spins maintains the initial state as the squeezing transfer channel between the two species is closed.  
		Next, a magnetic field $B_r=\eta (n_i-n_s)/(2\gamma_s-2\gamma_i)$ in
		the direction of collective spins is applied to atomic ensembles to offset the effective detuning $\eta (n_i-n_s)/2$ between alkali-metal and noble-gas spins and last for a duration of $\tau_{op} = \pi {J^{-1}}
		$. Thus the nuclear spins will evolve to the SSS with squeezing parameter $\xi^2_{s,0}$, which can be stored for a long time due to the long lifetime of noble-gas nuclear spins.  
		Similarly, for a SSS of a noble-gas nuclear spins to be measured, the SSS can be value-preservedly transferred to alkali-metal spins by applying a same magnetic field $B_r$. By measuring the squeezing of alkali-metal spins, the nuclear spin squeezing can be readout. 
	}
	\label{fig-S2}
\end{figure*}

In the main text, we propose the method of generating and measuring nuclear SSS by the squeezing transfer between alkali-metal and noble-gas spin ensembles.
Here the initial alkali-metal spin squeezing is required, and the corresponding studies on experimental implementation have been extensively reported. For instance, various schemes for the preparation of alkali-metal spin squeezing have been implemented based on quantum nondemolition (QND) measurement in hot alkali-metal atomic vapor~\cite{Bao2020_Spin, Kong2020_Measurement-induced, Kuzmich2000_Generation, Kuzmich1999_Quantum}. These experimental techniques and physical systems are ideal candidates for generating the initial alkali SSS. In addition, noble gases are commonly added as buffer gases to atomic vapor to reduce the diffusion rate of alkali-metal atoms and enhance the interaction between atoms and the laser. 
When no external magnetic field is applied to atomic ensembles, we can deduce from Eq.~(\ref{eq-S18}) that the squeezing parameter of noble-gas nuclear spins at the optimal evolution time is given by 
\begin{align}
	\xi_{i,op}^2 = \frac{4r \xi_{s,q}^2 +(r-1)^2}{(r+1)^2}    .
	\label{eq-S23}
\end{align} 
with the QND squeezing parameter $\xi_{s, q}^2 = 1/(1+\kappa^2)$. We see that the transfer efficiency of spin squeezing depends on the ratio $r$ of the atomic numbers of alkali and noble-gas spin ensembles. It reveals that if the difference in atomic numbers of two spin ensembles is large, the squeezing does not transfer between the alkali and noble-gas spins, as shown in the inset of Fig.~\ref{fig2}(c). Thus, the process of preparing the initial alkali-metal spin squeezing using the QND method can be implemented as usual after adding noble-gas atoms into the alkali-metal vapor cell. 
Even if the atomic numbers of the alkali-metal and noble-gas ensembles are close, the squeezing transfer can also be turned off by applying a magnetic field to increase the effective detuning $\Delta_s-\Delta_i$ between the two spin ensembles. When the preparation of the SSS of the alkali-metal spins is finished, a magnetic field in the direction of collective spins satisfying $B_r=\eta (n_i-n_s)/2(\gamma_s-\gamma_i)$ can be applied to the vapor cell to turn on the squeezing transfer between the alkali and noble-gas spins. According to the discussion in the main text, the noble-gas nuclear spins will evolve to a SSS with the same squeezing parameter as that of the initial alkali-metal spin squeezing. As demonstrated in Fig.~\ref{fig4}(a), the accuracy of the external magnetic field required in the above implementation is in the realizable parameter region experimentally. Therefore, the experimental technologies for the preparation of alkali-metal spin squeezing based on the QND method can be conveniently used for the preparation of nuclear spin squeezing by applying a magnetic field to control the squeezing transfer between the alkali and noble-gas spin ensembles.   

Based on the above discussion, the preparation, storage, and measurement of the nuclear spin squeezing can be implemented following the sequence shown in Fig.~\ref{fig-S2}. Specifically, once the electron spin squeezing of the alkali-metal atoms has been prepared, the magnetic field $B_r$ can be applied to transfer the spin squeezing to noble-gas nuclear spins. Subsequently, after a duration of $\tau_{op}$, we switch off the squeezing transfer process, and the electron spin squeezing is completely transferred to the nuclear spins and maintained stably, thus being stored in the nuclear spins. The significant advantage of the noble-gas nuclear spins is the long lifetime, which can exceed several days at or above room temperature due to the isolation of noble-gas nuclear spins from the environment, so that long-time storage of spin squeezing will be realized. 

However, the isolation nature of noble-gas nuclear spins also leads to a significant challenge in measuring nuclear spin squeezing at present. Fortunately, this problem can be solved in our protocol. To measure the squeezing degree of nuclear spin, the magnetic field $B_r$ is applied to turn on the squeezing transfer between the alkali and noble-gas spins, and then the time evolution of the squeezing parameter is given by 
\begin{eqnarray}
	\xi_{s,r}^2(t) &=\frac{1}{2} \left(1 + \xi_{i,m}^{2} + (1 - \xi_{i,m}^{2}) \cos(Jt)\right)   \nonumber   \\
	\xi_{i,r}^{2}(t) &= \frac{1}{2} \left(1 + \xi_{i,m}^{2} - (1 - \xi_{i,m}^{2}) \cos(Jt)\right)    .     
	\label{eq-S24}
\end{eqnarray} 
Here the $\xi_{i,m}^{2}$ denotes the squeezing parameter of the nuclear spins that needs to be measured.  
We can see that when the magnetic field $B_r$ is applied for a duration $\tau_{op}$, the nuclear spin squeezing to be measured will be transferred completely to the alkali-metal spins, whose spin squeezing can be measured through mature experimental technologies.

\section{Impact of imperfect polarization}
\label{App-4}

\begin{figure*}[htbp]
	\centering
	\includegraphics[width=18cm]{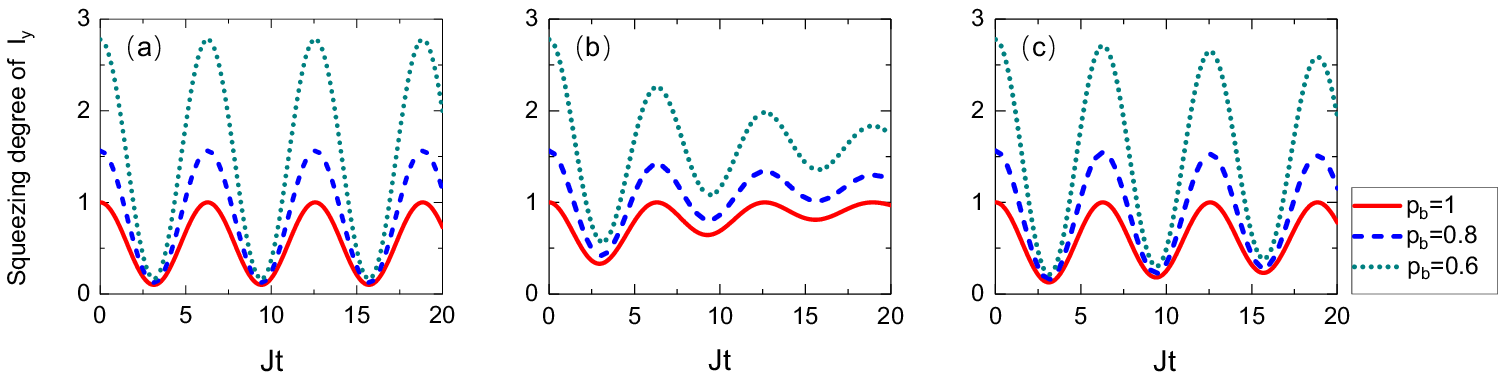}
	\caption{(Color online) 
		Evolution of the squeezing degree of the noble-gas nuclear spin component \(I_{y} \) with different polarizations.  Other parameters satisfy \(\Gamma_{s}=0\) in (a), \(J=10 \Gamma_{s}\) in (b), and \(J=100 \Gamma_{s}\) in (c). The initial alkali-metal spin squeezing degree is along x-axis and the squeezing degree satisfies \(\xi_{s, x}^{2}(0)=0.1\). 
	}
	\label{fig-S3}
\end{figure*}

Techniques for preparing highly polarized alkali-metal and noble-gas spins have been reported.
For instance, the polarization of alkali-metal spins can be almost close to 100\% in Ref.~\cite{Dideriksen2021_Room-temperature}. Based on spin-exchange optical pumping, the Helium-3 polarization of 85\% was achieved in Ref.~\cite{Chen2014_On}. 
Therefore, the preceding discussion neglected the impact of imperfect polarization in alkali-metal and noble-gas atoms.  
Next, to gain a comprehensive understanding, we will supplement the discussion on this impact.
For the mixed gases consisting of alkali-metal and noble-gas atoms, the dynamics of the spin operators considering imperfect polarization, relaxation process, and associated noise can be described by the explicit motion equations
\begin{align}
	\partial_{t} S^{-}=-\left(i \Delta_{s}+\frac{\Gamma_{s}}{2}\right) S^{-}-i \frac{J}{2} I^{-}+F_{s}, \nonumber   \\
	\partial_{t} I^{-}=-\left(i \Delta_{i}+\frac{\Gamma_{i}}{2}\right) I^{-}-i \frac{J}{2} S^{-}+F_{i} .
   	\label{eq-S4-1}
\end{align}
Here \(S^{-}=\frac{1}{\sqrt{p_{s} n_{s}}} \sum_{a=1}^{n_{s}} s_{a}^{-}\) and \(I^{-}=\frac{1}{\sqrt{p_{i} n_{i}}} \sum_{b=1}^{n_{i}} i_{b}^{-}\) denote the collective spin operator of alkali-metal and noble-gas atoms, respectively. \(J=\eta \sqrt{p_{s} p_{i} n_{s} n_{i}}\) denotes the collective exchange rate of the coherent spin-exchange interaction considering imperfect polarization, with the polarizations \(p_{s}\) and \(p_{i}\) of alkali-metal spins and noble-gas nuclear spins, respectively. \(F_{s}\) and \(F_{i}\) denote the noise operators, with \(\langle F_{s}(t) F_{s}^{+}(t')\rangle=2 \Gamma_{s} \delta(t-t')\), \(\langle F_{i}(t) F_{i}^{+}(t')\rangle=2 \Gamma_{i} \delta(t-t')\). For the case of resonance, the effective detuning \(\Delta_{s}\) and \(\Delta_{i}\) satisfy \(\Delta_{s}=\Delta_{i}=0\).

According to the analysis for the case without relaxation process in the main text, when the initial spin squeezing of the alkali-metal ensemble is along the x-axis, the directions of the optimal squeezing of the alkali-metal and noble-gas nuclear spins remain consistent with the directions of the spin components \(S_{x}\) and  \(I_{y}\), respectively, in the evolution over time. By solving Eq.~(\ref{eq-S4-1}), we obtain that the evolution of the variance of the spin component \(I_{y} \) satisfies
\begin{widetext}
\begin{align}
	\mathrm{Var}\left[I_{y}\right] & =-\frac{1}{4 J_{s}^{2}} e^{-\left(i J_{s}+\Gamma_{s}\right) t}\left(i J_{s}-\Gamma_{s}+e^{i J_{s} t}\left(i J_{s}+\Gamma_{s}\right)\right)^{2} \mathrm{Var}\left[I_{y}(0)\right] \nonumber   \\
	& -\frac{J^{2}}{4 J_{s}^{2}} e^{-\left(i J_{s}+\Gamma_{s}\right) t}\left(e^{i J_{s} t}-1\right)^{2} \mathrm{Var}\left[S_{x}(0)\right]          \nonumber   \\
	& +\frac{1}{4 J_{s}^{2} \Gamma_{s}}e^{-\left(i J_{s}+\Gamma_{s}\right) t}\left(\left(i J_{s}+\Gamma_{s}\right) \Gamma_{s} e^{2 i J_{s} t}-2 J^{2} e^{i J_{s} t}+2 J_{s}^{2} e^{i\left(J_{s}+\Gamma_{s}\right) t}+\Gamma_{s}\left(\Gamma_{s}-i J_{s}\right)\right) \mathrm{Var}\left[S_{x}\right]_{F_{s}},
	\label{eq-S4-2}
\end{align}
\end{widetext}
with \(J_{s}^{2}=\sqrt{J^{2}-\Gamma_{s}^{2}}\). The relaxation rate of noble-gas spins is several orders of magnitude smaller than that of alkali spins, and thus is neglected. From Eq.~(\ref{eq-S4-2}), one can see that the noise origins of the noble-gas nuclear spins can be divided into three components. The first and second terms represent the contributions from the initial states of the noble-gas nuclear spins and the alkali-metal electronic spins, respectively. The third term represents the contribution of the fluctuating noise. Since the atoms in both the unpolarized state and the CSS are uncorrelated, the initial variance of the spin component is
\begin{widetext}
\begin{align}
 	\mathrm{Var}\left[I_{y}(0)\right]=\frac{1}{p_{i} n_{i}}\left(\sum_{b=1}^{p_{i} n_{i}} \mathrm{Var}\left[i_{b, y}^{p}(0)\right] 
 	+\sum_{b=1}^{\left(1-p_{i}\right) n_{i}} \mathrm{Var}\left[i_{b, y}^{u n p}(0)\right]\right)=\frac{1}{4 p_{i}},
 		\label{eq-S4-3}
\end{align}
\end{widetext}
for the noble-gas nuclear spins with spin 1/2, with \(i_{b, y}^{p}\) and \(i_{b, y}^{u n p}\) denoting the polarized and unpolarized spins, respectively. 

In Fig.~\ref{fig-S3}, we present the evolution of the squeezing degree of the spin component \(I_{y}\) when the partially polarized noble-gas nuclear spins couple with the alkali-metal spins initially in spin-squeezed states. 
In Fig.~\ref{fig-S3}(a), the case without relaxation process, i.e., \(\Gamma_{s}=0\), is shown. It reveals that although different spin polarizations correspond to different initial states, the squeezing degree achievable for the noble-gas nuclear spins at the optimal evolution time, i.e., \(\tau_{op}=\pi J^{-1}\), is not significantly affected. 
This performance stems from the fact that the noble-gas spins acquire the initial spin squeezing of the alkali-metal atoms completely via spin-exchange interaction and the variance in Eq.~(\ref{eq-S4-2}) retains only the second term at \(t=\pi J^{-1}\).

Furthermore, when incoherent processes are considered, the polarization of atomic spins modulates the exchange rate \(J=\eta \sqrt{p_{s} p_{i} n_{s} n_{i}}\), thereby altering the coupling-to-decay ratio \(J/\Gamma_{s}\). To gain an intuitive understanding of this scenario, we present the evolution curves of the squeezing degree of noble-gas nuclear spins considering dissipation and quantum fluctuation processes in Fig.~\ref{fig-S3}(b) and (c), with different polarizations and coupling-to-decay ratios. 
These results indicate that, despite the unavoidable effects of imperfect polarization, effective squeezing transfer can be maintained when the strong-coupling regime of the coherent spin-exchange interaction is reached. 
Fortunately, current experimental techniques have demonstrated that this strong spin-exchange interaction is achievable \cite{Shaham2022_Strong}.

%\bibliographystyle{apsrev4-2}
\bibliography{bib_NSSS}

\end{document}